\documentclass[journal]{IEEEtran}
\usepackage[pdftex]{graphicx}
\usepackage{amsmath}
\usepackage{amssymb}
\usepackage{cite}

\usepackage{url}
\usepackage{enumitem}
\usepackage{float}
\usepackage{wasysym}
\usepackage{multirow}
\usepackage{mathrsfs}
\usepackage{xcolor}
\usepackage{slashbox}

\newcommand{\RN}[1]{%
  \textup{\uppercase\expandafter{\romannumeral#1}}%
}

\begin{document}
\title{A Distributed and Resilient Bargaining Game for Weather-Predictive Microgrid Energy Cooperation}

\author{\IEEEauthorblockN{Lu An, Jie Duan, Mo-Yuen Chow, Alexandra Duel-Hallen}\\
\IEEEauthorblockA{Department of Electrical and Computer Engineering\\
North Carolina State University\\
Raleigh, NC 27695\\
Email: \{lan4,jduan3,chow,sasha\}@ncsu.edu}}

\maketitle
\thispagestyle{plain}
\pagestyle{plain}

\begin{abstract}
A bargaining game is investigated for cooperative energy management in microgrids. This game incorporates a fully distributed and realistic cooperative power scheduling algorithm (CoDES) as well as a distributed Nash Bargaining Solution (NBS)-based method of allocating the overall power bill resulting from CoDES. A novel weather-based stochastic renewable generation (RG) prediction method is incorporated in the power scheduling. We demonstrate the proposed game using a 4-user grid-connected microgrid model with diverse user demands, storage, and RG profiles and examine the effect of weather prediction on day-ahead power scheduling and cost/profit allocation. Finally, the impact of users' ambivalence about cooperation and /or dishonesty on the bargaining outcome is investigated, and it is shown that the proposed game is resilient to malicious users' attempts to avoid payment of their fair share of the overall bill. 

\end{abstract}

\IEEEpeerreviewmaketitle
\vspace{-0.1in}

\section{Introduction}
\IEEEPARstart{M}{icrogrids} are autonomous small-scale power grid systems that support a flexible, reliable, and efficient integration of renewable sources of energy, such as solar and wind, distributed energy storage devices (DESDs), and demand response \cite{el2014smart}. The optimal power scheduling of generation and DESD charging/discharging to minimize the overall cost of power purchasing from the main grid is vital to efficient microgrid energy management \cite{8451959,7501514}.
To improve the system-wide economic performance and robustness of microgrids, cooperative energy management methods were proposed in \cite{8451959,7501514,7613189}. Moreover, distributed operation without utilizing a control center is vital to preserve users' privacy and optimize cost-effectiveness of energy management \cite{7582434}.
In \cite{rahbari2016consensus}, the authors proposed a dynamic, multi-step, Cooperative, and Distributed Energy Scheduling (CoDES) algorithm, which integrates RG and energy storage, to minimize the overall cost of purchasing power from the main grid. In \cite{AnPES17}, the CoDES algorithm was extended to include bi-directional power trading (both purchasing and selling) with the main grid.

Most distributed power scheduling algorithms are designed under the assumption that all microgrid users are willing to cooperate to produce the optimal system-wide schedule. However, in practice, such cooperation is not assured. Microgrids usually consist of multiple agents, e.g. households, who trade power among themselves and with the main grid, and have their individual economic and social optimization objectives. For example, they might be ambivalent about cooperation with some of their neighbors or they might try to reduce their individual costs. Cooperative game theory \cite{han2012game} is suitable for resolving cooperation and competition among agents and thus has been utilized extensively for energy trading and economic dispatch in microgrids, e.g. \cite{8071018,chakraborty2015real,8226810,tushar2015canonical}. For example, coalitional games, which identify subsets of cooperating users, were investigated for networked microgrids in  \cite{chakraborty2015real,8226810,tushar2015canonical}. However, such games require either centralized implementation or extensive exchange of users' parameters and have high computational complexity. Moreover, in a coalitional game, it is often difficult to find a practical payoff allocation that lies in the core, leading to stability issues \cite{chakraborty2015real}.

On the other hand, bargaining games \cite{han2012game} address scenarios where the users are willing to cooperate provided their individual objectives are met. In bargaining games, the total cost of social optimization is distributed among the users according to a cost allocation method. Several approaches to fair cost allocation for cooperative games have been proposed in \cite{Avrachenkov2015265,ROSENTHAL201764,7890990}. To date, most cost allocation methods in power grid focused either on splitting the incurred cost \cite{beeler2014network} or the deviation from the users' aggregated contract after power consumption \cite{ghavidel2018incentive}.
To the best of our knowledge, only \cite{AnPES17} addressed cost allocation for distributed microgrid power scheduling. In \cite{AnPES17}, a bargaining game that employed a computationally efficient NBS cost allocation \cite{Avrachenkov2015265} was developed to allocate the overall predicted power bill. However, in \cite{AnPES17} all users were confident in their neighbors and were honest. These idealistic assumptions ensure successful bargaining and, thus, cooperation on power scheduling, but they are not suitable in practice, especially in a distributed power scheduling environment, where the users do not disclose their individual power profiles and costs.

In this paper, we extend the game of \cite{AnPES17} to include user's ambivalence about cooperation as well as malicious reduction of the reported selfish cost or profit for individual user's economic advantage. In this case, bargaining might fail, thus preventing the users from cooperating on power scheduling. We analyze the dependency of the bargaining success on individual users' power profiles as well as their ambivalence and dishonesty. Moreover, we demonstrate that the proposed game is resilient to malicious selfish costs reduction. The proposed NBS cost allocation is computationally simple and is computed using the averaging consensus algorithm \cite{4118472}, thus resulting in distributed implementation.

The objective of cost allocation in this paper is to aid the cooperative power scheduling for an upcoming time interval, e.g., a day ahead. It provides users with incentives to cooperate or to reassess the cooperating agreement. 
If the bargaining outcome is positive, the users' allocated costs can be viewed as their individual contracts with the utility company, which are augmented after consumption if the actual power cost deviates from its predicted value due to the demand, RG, and price variation, using e.g. \cite{beeler2014network,ghavidel2018incentive}. If bargaining fails, users might choose to act individually or make another cooperation attempt. Finally, the communication cost of the power scheduling and cost allocation algorithms \cite{lian2016game}, real-time power scheduling updates, expenses of microgrid system security, and other costs associated with user cooperation must also be fairly allocated. While these issues can in principle be incorporated into the proposed bargaining game, they are beyond the scope of this work.

In this paper, we also introduce utilization of \textit{weather-based RG prediction} in power scheduling. Most distributed power scheduling algorithms adopt deterministic RG profiles \cite{ 7582434,rahbari2016consensus,AnPES17}, which are not realistic due to weather uncertainty. To capture this uncertainty, \cite{7154496} modeled RG using exact known distributions, and \cite{7544637} proposed a stochastic energy scheduling model where the forecasting error of RG was modeled by arbitrary realizations over time. Moreover, \cite{stochasticES2014} proposed a scenarios-based stochastic RG model, with the RG profile for the next scheduling interval chosen equiprobably from the scenario pool. However, energy management approaches in the literature did not take into account the weather forecast, which can greatly impact predicted solar and wind RG, as was shown in other smart grid applications, e.g. \cite{7459241,6802349}. We propose to include a weather-based stochastic RG prediction method into the proposed energy management game. Finally, the battery degradation costs (BDC) \cite{7744978,wu2018energy} are also taken into account to aid more realistic power scheduling.

The main \textit{contributions} of this paper are:
\begin{itemize}[leftmargin=*]
\item Formulation of a bargaining game that includes a fair, computationally-efficient, and fully distributed cost-allocation algorithm for predictive energy management in cooperative microgrids.

\item Analysis of the impacts of ambivalent and dishonest users on the bargaining outcome and demonstration of resilience of the proposed game to selfish cost manipulation.

\item Inclusion of weather-based stochastic RG prediction into energy management and analysis of the proposed game for realistic microgrids.

\end{itemize}

The rest of the paper is organized as follows. Section II presents the microgrid system model and the bargaining game. In Section III, the cooperative and individual power scheduling optimization problems are formulated, and the proposed weather-based stochastic RG prediction method is described. The distributed NBS-based cost allocation algorithm is contained in Section IV. This section also discusses the impact of ambivalent and dishonest users on the bargaining outcome. Numerical results for a heterogeneous 4-user microgrid are contained in Section V. Section VI concludes the paper.

\vspace{-0.1in}
\section{Microgrid system model and the bargaining game} \label{sysmod}
\begin{table*}[h]
\centering
\caption{Frequently Used Notation}
\label{tab:note}
\vspace{-0.1in}
\begin{tabular}{|p{15mm}|p{60mm}||p{15mm}|p{60mm}|}
\hline
\textbf{Term }& \textbf{Definition}       & \textbf{Term} & \textbf{Definition} \\ \hline
$U_{act}$/$U_{pas}$ & The sets of active/passive users   & $U_{PV}$/$U_{WT}$ & The sets of active users who own PV/WT generators \\ \hline
 $\Delta t$              & Time interval between two time steps (hr)      & $T$/$t$           & Number of time steps for power scheduling/current time step \\ \hline
  $r$              & The total number of users      & $i$           & The index of user \\ \hline
$p_b (t)$/$p_s(t)$              &   Prices at which the users purchase/sell power from/to the grid (given profile)                                & $\kappa_i$               & Discharging/charging efficiency \cite{7744978} of DESD $i$ \\ \hline
$P_G^+ (t)\geq 0$               & Power purchased from the grid                                                          &  $P_G^- (t)\geq 0$            &   Power sold to the grid                                                           \\ \hline
$P_{i,B}^+ (t)\geq 0$            &   The discharging amount of DESD $i$ & $P_{i,B}^- (t)\geq 0$               & The charging amount of DESD $i$ \\ \hline
$P_G(t)$               & ${ {P_G^ + \left( t \right) }  - P_G^ - \left( t \right)}$ \newline The net power obtained from the grid & $P_{i,B}(t)$            &   $P_{i,B}^+ (t)-P_{i,B}^- (t)$ \newline The net power command of DESD $i$  \\ \hline
$P_{i,D}(t)$            &   The forecasted load demand of user $i$ & $P_{i,R}(t)$            &   The predicted RG of user $i$ \\ \hline
$E_{i,B}^0$            &   The initial energy stored in DESD $i$ & $E_{i,B}^{\min}$/$E_{i,B}^{\max}$           &   The lower/upper capacity limits of DESD $i$ \\ \hline
$P_G^{\max}$            &   The upper bound on the power $P_G^+ (t)$/$P_G^- (t)$ that can be purchased from or sold to the grid & $P_{i,B}^{\max}$          &   The upper bound on the power $P_{i,B}^+ (t)$/$P_{i,B}^- (t)$ that DESD $i$ can discharge or charge \\ \hline
$C^{\mathcal{P}}_{opt}(t)$            &   The optimal power trading cost & $C^{i,\mathcal{B}}_{opt}(t)$          &   The optimal individual BDC of active user $i$ \\ \hline
$J_{soc}$            &   The social cost & $D_i$ & The ideal selfish cost of user $i$ \\ \hline
$S_i$            &   The selfish cost of user $i$ & $J_i$          & The allocated cost of user $i$ \\ \hline
$\gamma_i$            &   The selfish cost adjustment factor of user $i$ & $\epsilon$          & The cooperation discount \\ \hline
\end{tabular}
\vspace{-0.2in}
\end{table*}

\begin{figure}[h]
	\centering
	\vspace{-0.2in}
    \includegraphics[width=0.4\textwidth]{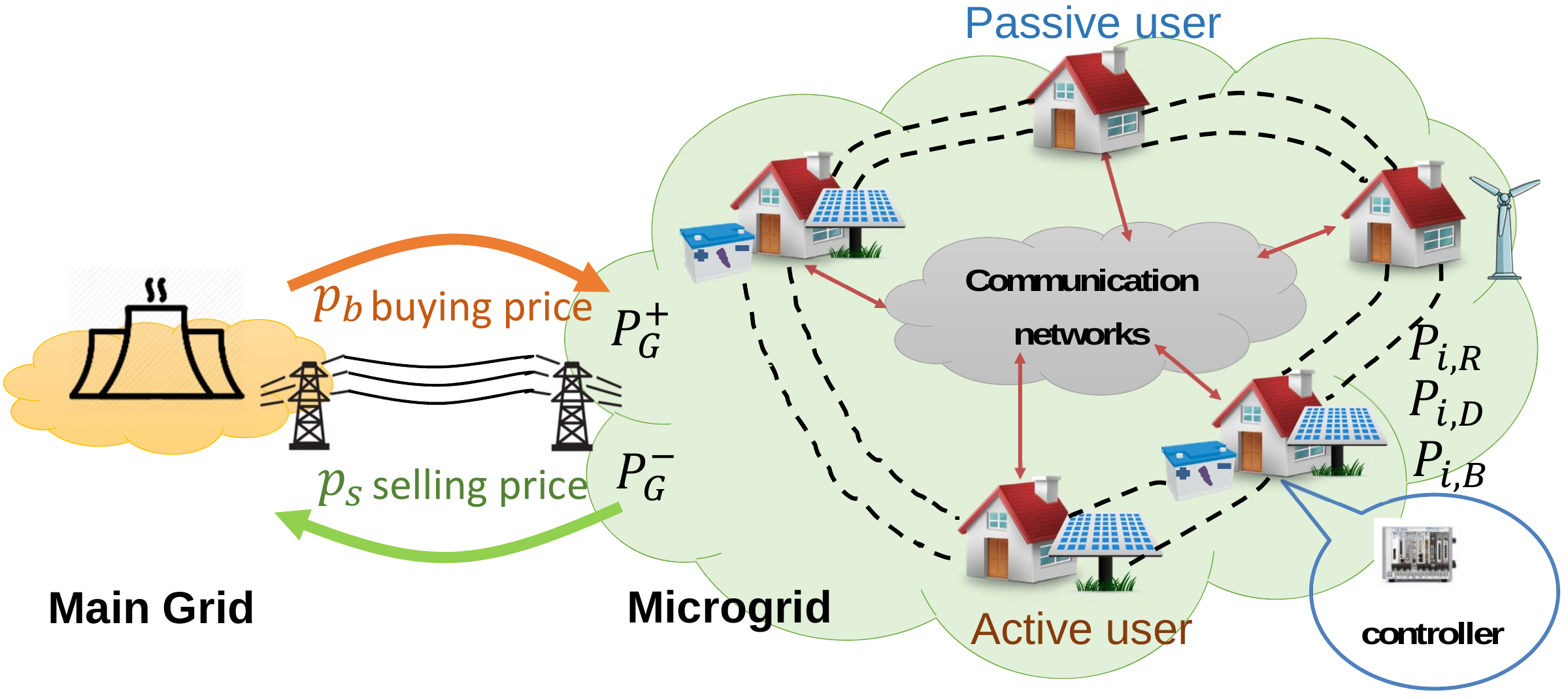}
    \vspace{-0.1in}
    \caption{Grid-connected microgrid system}
    \label{fig:mg}
\end{figure}

Consider a grid-connected microgrid system shown in Fig. \ref{fig:mg}. The supply side is the main grid, and the demand side (the microgrid) has $r$ users, including both \textit{passive} users, which own only loads, and \textit{active} users, which also own DESDs and/or RG units, e.g., wind turbines (WT) or photovoltaics (PV) panels. The passive users are energy consumers while the active users can both produce and consume energy. In a smart microgrid, distributed controllers are embedded in physical devices, enabling the users to communicate with their neighbors via bi-directional communication links, execute the designated algorithms, and optimize power scheduling of the local devices. We assume each user is interested in cooperating on power scheduling to minimize the overall power bill from the main grid, provided its individual economic and social objectives are satisfied. Table \ref{tab:note} lists the notation frequently used in the paper.
    \vspace{-0.1in}

To investigate the trade-offs of cooperation in power scheduling, we develop a bargaining game with microgrid users as players, who employ a computationally efficient NBS cost allocation method \cite{Avrachenkov2015265} to split the overall cost of the power bill obtained using cooperative power scheduling. 

For a system with $r$ players, the NBS cost allocation algorithm proceeds in \textit{three steps} \cite{Avrachenkov2015265}:

(1) The players jointly minimize the \textit{social cost} $J_{soc}$, i.e. the system-wide cost when all players cooperate.

(2) The \textit{disagreement point} is computed as $\mathbf{S}=(S_1,S_2,\cdots, S_{r})$, where the \textit{selfish cost} $S_i$ is the maximum cost the $i^{\text{th}}$ player is willing to pay.

(3) The overall cost $J_{soc}$ is split among the players, with the \textit{allocated cost} of player $i$ given by [Theorem 2, \cite{Avrachenkov2015265}]:
\begin{equation}
\label{eq:Ji}
{J_i} = {S_i} - \frac{{\sum\nolimits_{i = 1}^{r} {{S_i}}}  - J_{soc}}{r} \quad \forall i = 1, \cdots ,r.
\vspace{-0.1in}
\end{equation}
Note that $S_i>0$ or $J_i>0$ corresponds to the player $i$'s expenditure while $S_i<0$ or $J_i<0$ implies player $i$ is compensated for selling power and for cooperation. Moreover, bargaining is successful when the social cost does not exceed the sum of the selfish costs, i.e.,
\begin{equation}
\label{eq:JlessD}
J_{soc}\leq \sum\nolimits_{i = 1}^r {{S_i}},
\vspace{-0.1in}
\end{equation}
or, equivalently, each player's allocated cost is no greater than its selfish cost:
\vspace{-0.1in}
\begin{equation}
\vspace{-0.1in}
\label{eq:eps}
\epsilon = S_i - J_i \geq 0,
\end{equation}
where $ \epsilon$ is the discount of cooperation, which is the same for all players. Note that $\epsilon$ represents profit due to cooperation for compensated users. In the rest of the paper, we will refer to $S_i$ and $J_i$ as costs and to $\epsilon$ as discount for both charged and compensated users. By Theorem 2 in \cite{Avrachenkov2015265}, the cost $J_i$ in (\ref{eq:Ji}) is the smallest achievable allocated cost of player $i$ given $J_{soc}$ and $\mathbf{S}$ in steps 1 and 2, respectively, thus resulting in a fair and unique cost allocation.

Finally, we require \textit{distributed} implementation of the proposed cooperative game to eliminate the need for a central controller and to maintain privacy of individual users. Thus, the users will not exchange any information about their individual resources or power demands and will minimize other private information sharing. In the rest of the paper, we describe the steps of the proposed game and demonstrate that it effectively tests feasibility of cooperation for realistic microgrids and discourages dishonest users while maintaining user privacy and fast convergence.

\vspace{-0.1in}

\section{Power scheduling and weather-based renewable generation prediction}

\subsection{Cooperative Power Scheduling}
First, we compute the social cost $J_{soc}$ in step 1 of NBS in Section II. 
Assuming all demand-side users collaborate to minimize the power bill from the utility company for the upcoming time interval of $T$ time steps, the optimization problem can be described as:
\vspace{-0.1in}
\begin{equation}
\label{eq:min_J}
J_{soc} = \mathop {\min }\limits_{{{\mathbf{P}}}(t)} {\sum_{t = 1}^T {\left( {{C^\mathcal{P}}(t) + \sum\limits_{i \in {U_{act}}} {{C^{i,\mathcal{B}}}(t)}} \right)\Delta t}} ,
\vspace{-0.1in}
\end{equation}
where $C^\mathcal{P}$ is the power trading cost given by:
\begin{equation}
\label{eq:C_P}
{C^\mathcal{P}}(t) = {p_b}(t)P_G^ + \left( t \right) - {p_s}(t)P_G^ - \left( t \right).
\end{equation}
and $C^{i,\mathcal{B}}$ is the individual BDC \cite{7744978}:
\begin{equation}
\label{eq:C_iB}
{C^{i,\mathcal{B}}}(t) = {C_{i,d}}(t)\left( {P_{i,B}^ + (t) + P_{i,B}^ - (t)} \right),
\end{equation}
where $C_{i,d}(t)$ is the unit BDC of DESD $i$ defined in \cite{7744978}.

The total bill in (\ref{eq:min_J}) is minimized by optimizing the microgrid power schedule ${{\mathbf{P}}}(t)$ for the grid and all users $i=1,\cdots,r$, where at time step $t=1,\cdots,T$:
\vspace{-0.1in}
\begin{equation}
\label{eq:P}
\begin{gathered}
  {\mathbf{P}}(t) = [P_G^ + (t),P_G^ - (t),P_{1,B}^ + (t),P_{1,B}^ - (t), \cdots ,P_{r,B}^ + (t),P_{r,B}^ - (t)]. \hfill \\ 
\end{gathered}
\end{equation}

The optimization problem (\ref{eq:min_J}) is subject to the following constraints:
\begin{enumerate}[wide, labelwidth=!, labelindent=0pt]
\item \textit{Power Balance Constraint}: At any time step $t$, the amount of load is equal to the amount of power generation: 
\vspace{-0.1in}
\begin{equation}
\label{eq:power_balance}
\begin{gathered}
  \sum\limits_{i = 1}^n {{P_{i,D}}\left( t \right)} = 
   \sum\limits_{i = 1}^n {\left( {P_{i,B}^ + (t) - P_{i,B}^ - (t) + {P_{i,R}}(t)} \right)}  + {P_G}(t). \hfill \\ 
\end{gathered}
\end{equation}

\item \textit{DESD Dynamics and Capacity Limits}: The system states are given by the current stored energy in DESD $i$, which is limited by the battery capacity:
\begin{equation}
\label{eq:DESD_limit}
E_{i,B}^{\min } \leqslant E_{i,B}^0 - \sum\limits_{\tau  = 1}^t {\left({\frac{{P_{i,B}^ + \left( \tau \right)}}{{{\kappa _i}}}}-{{\kappa _i}P_{i,B}^ - \left( \tau \right)}\right)\Delta t}  \leqslant E_{i,B}^{\max }.
\end{equation}

\item \textit{Power Rating Constraints}: At any time step, the discharging/charging power of DESDs and the power traded with the grid must be in the range of the physical limits:
\vspace{-0.1in}
\begin{equation}
\label{eq:PiB_lim}
0 \leqslant P_{i,B}^ + \left( t \right),P_{i,B}^ - \left( t \right) \leqslant P_{i,B}^{\max },
\vspace{-0.1in}
\end{equation}
\begin{equation}
\label{eq:PG_lim}
0 \leqslant P_{G}^ + \left( t \right),P_{G}^ - \left( t \right) \leqslant P_{G}^{\max }.
\vspace{-0.1in}
\end{equation}
\end{enumerate}

The optimization process is summarized as follows. The inputs include stochastic RG predictions obtained from the weather-based scenario approach in Sec.III.C, the demands predicted by the users for the next time interval $T$, and the pricing profiles given by utility company. Then, the optimization problem (\ref{eq:min_J}) is solved by modifying the CoDES algorithm in \cite{AnPES17} with the Lagrangian function adjusted according to the cost function (\ref{eq:min_J}) and the constraints (\ref{eq:power_balance})-(\ref{eq:PG_lim}), which take into account the BDC factor. The update and consensus process is similar to that in CoDES \cite{AnPES17}. The outputs are the optimal power schedule ${{\mathbf{P}}^*}(t)$ for the grid and the DESDs, the optimal power trading cost ${C_{opt}^\mathcal{P}} = \sum_{t = 1}^T {{C^\mathcal{P}}(t)} \Delta t$ and the individual BDC ${C_{opt}^{i,\mathcal{B}}}=\sum_{t = 1}^T {{C^{i,\mathcal{B}}}(t)} \Delta t$ over the scheduling period, and the social cost $J_{soc}=C_{opt}^\mathcal{P}+\sum_{{i \in {U_{act}}}} {C_{opt}^{i,\mathcal{B}}}$, used in step 1 of the game.

Note that the CoDES algorithm satisfies the requirement of user privacy and fully distributed implementation for the proposed game \cite{rahbari2016consensus}. In particular, the power trading cost ${C_{opt}^\mathcal{P}}$ is a function of grid's variables $[{P_G^ {+*} (t)},{P_G^ {-*} (t)}]$ and thus is a private value of the grid while the individual battery degradation cost ${C_{opt}^{i,\mathcal{B}}}$ is a function of $[{P_{i,B}^ {+*} (t)},{P_{i,B}^ {-*} (t)}]$ and thus is a private value of the active user $i$. Moreover, the knowledge of the total cost $J_{soc}$ is not available to the grid or the users. These features will be utilized in the distributed cost allocation method in Sec.IV.B.
\vspace{-0.1in}
\subsection{Individual Power Scheduling}
\label{sec:ca}
To compute the selfish cost in step 2 of the bargaining game, each player first needs to estimate the cost of trading with the grid individually, i.e. when not cooperating with other users. For the $i^{\text{th}}$ player, denote the power drawn from and injected into the grid as $P_{i,G}^ + \left( t \right)$ and $P_{i,G}^ - \left( t \right)$, respectively. Then the following constraints must be satisfied:
\begin{equation}
\label{eq:ith_PG+_cons}
0 \leqslant P_{i,G}^ + \left( t \right) \leqslant P_G^{\max }, \quad 0 \leqslant P_{i,G}^ - \left( t \right) \leqslant P_G^{\max }.
\end{equation}
Thus, the $i^{\text{th}}$ player's individual power trading cost can be written as:
\vspace{-0.1in}
\begin{equation}
\label{eq:C_iP}
{C^{i,\mathcal{P}}}(t) = {p_b}(t){P_{i,G}^ + \left( t \right) }  - {p_s}(t)P_{i,G}^ - \left( t \right).
\end{equation}

Similarly to (\ref{eq:power_balance}), the power balance constraint for player $i=1,\cdots,r$ must be satisfied as:
\vspace{-0.1in}
\begin{equation}
\label{eq:ithpower_balance}
{P_{i,D}}\left( t \right) = \left( {P_{i,B}^ + (t) - P_{i,B}^ - (t) + {P_{i,R}}(t) } \right) + P_{i,G}(t),
\vspace{-0.1in}
\end{equation}
where  $P_{i,G}(t)={ {P_{i,G}^ + \left( t \right)} - P_{i,G}^ - \left( t \right)}$ is the net power that the $i^{\text{th}}$ user obtains from the grid. 
Note that (\ref{eq:ithpower_balance}) also applies to passive users by setting $P_{i,B}^ + (t)=P_{i,B}^ - (t)={P_{i,R}}(t)=0$. The individual BDC of active user $i$ is given by (\ref{eq:C_iB}).

Denote the $i^{\text{th}}$ user's individual power rating commands as:
\begin{equation}
{{\mathbf{P}}_{\mathbf{i}}}(t) = \left\{ {\begin{array}{*{20}{c}}
  {\left[ {\begin{array}{*{10}{c}}
  {P_{i,G}^ + (t)}&{P_{i,G}^ - (t)} 
\end{array}} \right],\quad i \in {U_{pas}}} \\ 
  {\left[ {\begin{array}{*{10}{c}}
  {P_{i,G}^ + (t)}&{P_{i,G}^ - (t)}&{P_{i,B}^ + (t)}&{P_{i,B}^ - (t)} 
\end{array}} \right], i \in {U_{act}}} 
\end{array}} \right. .
\end{equation}
Similarly to (\ref{eq:min_J}), we can express user $i$'s individual cost of power trading and battery degradation as:
\vspace{-0.1in}
\begin{equation}
\label{eq:Di}
{D_i} = \mathop {\min }\limits_{{{\mathbf{P}}_{\mathbf{i}}}(t)} \sum_{t = 1}^T {\left( {{C^{i,P}}(t) + {C^{i,\mathcal{B}}}(t)} \right)} \Delta t,
\vspace{-0.1in}
\end{equation}
which is subject to the individual power balance constraints (\ref{eq:ithpower_balance}), battery capacity limitation (\ref{eq:DESD_limit}), and power rating constraints (\ref{eq:PiB_lim}) and (\ref{eq:ith_PG+_cons}). Note that (\ref{eq:Di}) can be solved by user $i$ individually using nonlinear programming \cite{bazaraa2013nonlinear} and does not require data from other users.

By solving (\ref{eq:Di}), each user computes the smallest power bill $D_i$ it can achieve when not cooperating with other users. If $D_i$ is negative, the active user $i$ expects the profit $|D_i|$ by selling its generated power to the main grid. The cost $D_i$ will be employed in computing the disagreement point $\mathbf{S}$ as described in Sec.IV.A.

\vspace{-0.2in}
\subsection{Weather-based Renewable Generation Prediction}
\label{sec:wbr}
The energy management methods in Sec. III.A and III.B require accurate prediction of renewable power generation for the upcoming time period $T$, e.g. a day ahead. Since such generation is highly dependent on weather conditions, we employ weather-based stochastic renewable power prediction. We utilize a scenario pool \cite{stochasticES2014} to choose a suitable predicted RG power profile. However, in \cite{stochasticES2014}, the scenarios are equally likely, while in our approach, the probabilities of the scenarios are determined by the weather forecast for the upcoming time interval $T$.
\begin{figure}[h]
	\centering
	\vspace{-0.1in}
    \includegraphics[width=0.4\textwidth]{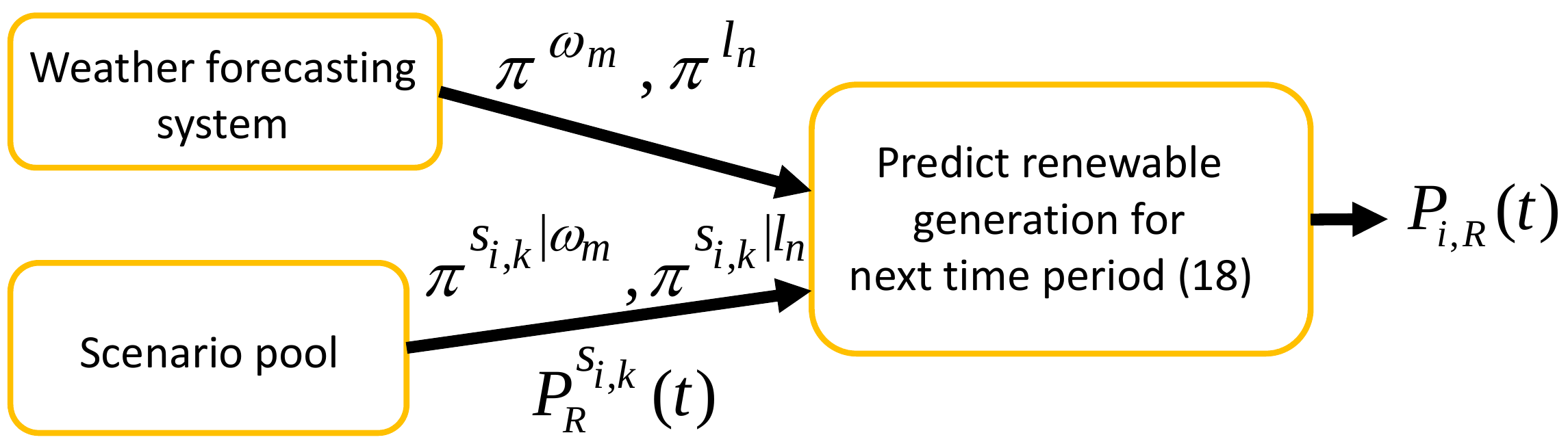}
    \vspace{-0.1in}
    \caption{Weather-based stochastic renewable generation prediction}
    \label{fig:WBR}
        \vspace{-0.1in}
\end{figure}

Fig. \ref{fig:WBR} illustrates the proposed method. A weather forecasting system predicts the weather type and is characterized by independent random variables $W$ and $L$, with the sample spaces ${\mathbf{\Omega}}=\{\text{Sunny: }\omega_1, \text{Cloudy: }\omega_2, \text{Rainy: }\omega_3\}$ and $\mathbf{L}=\{\text{Level 1: }l_1, \text{Level 2: }l_2, \text{Level 3: }l_3, \text{Level 4: }l_4\}$, respectively (other sample spaces can be chosen). The value $\omega_m\in {\mathbf{\Omega}}$ of $W$ provides the forecast for PV users for the time period $T$, e.g. for the next day, with the probability $P(W=\omega_m)=\pi^{\omega_m},m=1,2,3$. Similarly, for WT users, the wind speed level's value is given by $l_n\in \mathbf{L}$, with the probability $P(L=l_n)=\pi^{l_n}, n=1,2,3,4$. For example, if the forecasting system predicts that the next day is sunny/cloudy with probabilities 0.9 and 0.1, respectively, then $\pi^{\omega_1}=0.9$, $\pi^{\omega_2}=0.1$, and $\pi^{\omega_3}=0$.

Moreover, user $i$ employs a scenario pool, with the $k^{\text{th}}$ scenario denoted $s_{i,k}$ for $k=1,\cdots, K$, where $s_{i,k}$ corresponds to a certain RG profile of user $i$, $P_{R}^{s_{i,k}}(t), \forall i$, for $t=1,\cdots, T$. Given weather types $\omega_m$ and $l_n$, the probabilities that the scenario $s_{i,k}$ will occur are denoted by $\pi^{s_{i,k}|\omega_m}$ and $\pi^{s_{i,k}|l_n}$ for the PV and the WT owners, respectively.

Thus, with the weather forecast results $\pi^{\omega_m}$ and $\pi^{l_n}$, we can express the expected renewable generated power at the time unit $t=1,\cdots,T$ as:
\begin{equation}
\label{eq:wbr}
{P_{i,R}}(t) = \left\{ {\begin{array}{*{20}{c}}
  {\sum\limits_{m = 1}^3 {\pi^{{\omega _m}}\sum\limits_{k = 1}^K {\pi^{{s_{i,k}}|{\omega _m}}P_{R}^{{s_{i,k}}}\left( t \right),} } }&{i \in {U_{PV}}} \\ 
  {\sum\limits_{n = 1}^4 {\pi^{{l_n}}\sum\limits_{k = 1}^K {\pi^{{s_{i,k}}|{l_n}}P_{R}^{{s_{i,k}}}\left( t \right),} } }&{i \in {U_{WT}}} 
\end{array}} \right..
\end{equation}

The weather-based stochastic RG prediction $P_{i,R}(t)$, for $t=1,\cdots,T$ is used as an input in cooperative (\ref{eq:min_J}) and individual (\ref{eq:Di}) power scheduling problems in Sec.III.A and III.B, respectively.

\vspace{-0.1in}
\section{Distributed and resilient cost allocation to enable realistic energy management}
\label{sec:CA}
\subsection{The Selfish Cost}
\label{sec:dca}

As discussed in Sec.III.A and B, the users jointly compute the social cost $J_{soc}$ (\ref{eq:min_J}) in step 1 of the bargaining game in Sec.II, and each user $i$ computes its optimal individual power bill $D_i$ (\ref{eq:Di}), which will be referred to as the \textit{ideal} selfish cost. Next, user $i$ determines its actual selfish cost $S_i$ in step 2 of the game based on $D_i$. Given $J_{soc}$, if bargaining is successful, i.e. if (\ref{eq:JlessD}) holds, then eq. (\ref{eq:Ji}) implies that a lower selfish cost $S_i$ results in a lower allocated cost $J_i$ for user $i$. Thus, a user might want to report a lower selfish cost (or a greater profit) than $D_i$ if it is ambivalent about potential security and privacy problems of cooperation or about integrity of some of its neighbors. The corresponding reduction of that user's contribution to the overall bill might compensate for the user's ambivalence and entice it to cooperate. Moreover, a dishonest user might try to gain economic benefit (lower its allocated cost) by reducing its reported power cost. Define the $i^{\text{th}}$ user's selfish cost as:
\vspace{-0.1in}
\begin{equation}
\label{eq:Si}
{S_i} = {D_i} - {\gamma _i}\left| {{D_i}} \right|,
\vspace{-0.1in}
\end{equation}
where $\gamma_i \geq 0$ is the selfish cost adjustment factor of the $i^{\text{th}}$ user, which is the product of the ambivalence and dishonesty factors, and $\gamma_i|D_i|$ is the $i^{\text{th}}$ user's selfish cost reduction (SCR). Note that $\gamma_i=0$ or $S_i=D_i$ indicates the user's lack of ambivalence and honesty while large $\gamma_i$ means this user is very hesitant about the cooperation and/or extremely dishonest.

\vspace{-0.2in}
\subsection{Distributed Cost Allocation}
Using $J_{soc}$ and $S_i$ values, the cost allocation can be computed as in (\ref{eq:Ji}). However, in (\ref{eq:Ji}), each agent needs the knowledge of other player's selfish costs $S_j, \forall j\neq i$ and the social cost $J_{soc}$, which are the private infomation of the other players and the grid, repectively. To protect users' privacy and reduce the exchange of these costs, we employ the averaging consensus algorithm \cite{4118472} where the players and the grid communicate only with their neighbors. Set the initial states of the players as:
\vspace{-0.1in}
\begin{equation}
\label{eq:x0}
{x_i}(0) = \left\{ {\begin{array}{*{20}{c}}
  {{S_i},}&{i \in {U_{pas}}} \\ 
  {{S_i} - C_{opt}^{i,\mathcal{B}},}&{i \in {U_{act}}} 
\end{array}} \right.,
\vspace{-0.1in}
\end{equation}
and set the grid's initial state to ${x_{r + 1}}(0) =  - C_{opt}^\mathcal{P}$. The state update equation of each node $i=1,\cdots,r+1$ is given by:
\vspace{-0.1in}
\begin{equation}
\label{eq:xi_up}
{x_i}(k + 1) = {x_i}(k) + \sum\nolimits_{j \in {N_i}}^{} {{\alpha_{ij}}\left( {{x_j}(k) - {x_i}(k)} \right)}.
\end{equation}
where $\alpha_{ij}$ is the entry $(i,j)$ of a doubly-stochastic consensus update matrix $\textbf{A}$, which represents the communication connectivity strength between node $i$ and $j$ \cite{4118472}, and $N_i$ is the set of neighbors of node $i$. Provided the communication network is connected, the state information $x_i(k)$ converges to the average of the initial states \cite{4118472}: 
\vspace{-0.1in}
\begin{equation}
\label{eq:x_avg}
  {{\hat x}_i} = \mathop {\lim }\limits_{k \to \infty } {x_i}(k) = \frac{{\sum\nolimits_{i = 1}^{r + 1} {{x_i}(0)} }}{{r + 1}} = \frac{{\sum\limits_{i = 1}^r {{S_i}}  - J_{soc}}}{{r + 1}}.
\end{equation}
At convergence, each agent is able to calculate its own allocated cost $J_i$ (\ref{eq:Ji}) using only its local information:
\begin{equation}
\label{eq:Ji_dis}
{J_i} = {S_i} - \frac{{{(r+1)}{{\hat x_i}}}}{{r}}, \quad \forall i = 1, \cdots ,r,
\end{equation}
which results in distributed cost allocation.
\vspace{-0.1in}
\subsection{Bargaining Outcome and Resilience to Dishonest Cost Reporting}
When the ideal selfish costs $D_i$ (\ref{eq:Di}) are reported by all users, i.e. all users are confident about cooperation and are honest, the resulting ideal allocated cost \cite{AnPES17} of user $i$ is:
\begin{equation}
\label{eq:Ji0}
J_i^0 = {D_i} - \frac{{\sum_{i = 1}^r {{D_i}}  - J_{soc}}}{r}.
\end{equation}
In the ideal case (\ref{eq:Ji0}), bargaining always holds, i.e. ${\sum_{i = 1}^r {{D_i}} \geq J_{soc}}$ because the set of selfish power schedules computed using (\ref{eq:Di}) for all agents is a suboptimal solution of the social optimization (\ref{eq:min_J}). In this case, each user receives the same ideal discount (\ref{eq:eps}):
\begin{equation}
\label{eq:eps0}
\epsilon^0=\frac{{\sum_{i = 1}^r {{D_i}}  - J_{soc}}}{r}.
\vspace{-0.1in}
\end{equation}
Similarly for the ideal case when $S_i=D_i$, it can easily be shown that any arrangement of such users into cooperating groups (coalitions) results in a bill at least as large as $J_{soc}$.

On the other hand, in a realistic case, when some adjustment factors $\gamma_i$ are positive due to users' ambivalence and/or dishonesty, bargaining might fail. By substituting $S_i$ (\ref{eq:Si}) into (\ref{eq:Ji}), the allocated cost of the $i^{\text{th}}$ user can be expressed as:
\begin{equation}
\label{eq:newJi}
{J_i} = {D_i} - {\gamma _i}\left| {{D_i}} \right| - \frac{{r{\epsilon ^0} - \sum_{i=1}^r {{\gamma _i}\left| {{D_i}} \right|} }}{r}.
\end{equation}
From (\ref{eq:newJi}), bargaining fails when the total ideal discount $r\epsilon^0$ is not sufficient to compensate for the total SCR:
\begin{equation}
\label{eq:SCR}
{R_{tot}} = \sum\nolimits_{i = 1}^r {{\gamma _i}\left| {{D_i}} \right|} .
\end{equation}
Thus, the condition for successful bargaining is:
\begin{equation}
\label{eq:gi_cond}
{R_{tot}} \leq r{\epsilon ^0}.
\vspace{-0.1in}
\end{equation}
Let $\sigma_i$ denote the SCR for all users except user $i$:
\begin{equation}
\label{eq:sig}
{\sigma _i} = \sum\nolimits_{j = 1,j \ne i}^r {{\gamma _j}\left| {{D_j}} \right|} .
\end{equation}
Then the total SCR (\ref{eq:SCR}) can be rewritten as:
\begin{equation}
\label{eq:Rtot}
R_{tot}=\gamma_i|D_i|+\sigma_i,
\vspace{-0.1in}
\end{equation}
and from (\ref{eq:gi_cond}), cooperation holds when
\begin{equation}
\label{eq:upb}
{\gamma _i} \leq \frac{{r{\epsilon ^0} - \sigma_i }}{{|{D_i}|}}.
\vspace{-0.1in}
\end{equation}

Moreover, when a \textit{dishonest} user $i$ considers manipulation of its power cost $D_i$, the ultimate goal is to reduce the resulting allocated cost $J_i$ relative to the case when that user acts honestly (\ref{eq:Ji0}). For example, assume all users are not ambivalent. Then from (\ref{eq:Ji}), (\ref{eq:Si}), (\ref{eq:Ji0}) and (\ref{eq:SCR}), the dishonest user's allocated cost can be expressed as:
\begin{equation}
\label{eq:Ji_mal}
\begin{gathered}
  {J_i} = {D_i} - {\gamma _i}\left| {{D_i}} \right| - \frac{{\sum_i^r {{D_i}}  - \sum_i^r {{\gamma _i}\left| {{D_i}} \right|}  - {J_{soc}}}}{r} \hfill \\
   = {D_i} - \frac{{\sum_i^r {{D_i}}  - {J_{soc}}}}{r} - \left( {{\gamma _i}\left| {{D_i}} \right| - \frac{{\sum_i^r {{\gamma _i}\left| {{D_i}} \right|} }}{r}} \right) \hfill \\
   = J_i^0 - \left( {{\gamma _i}\left| {{D_i}} \right| - \frac{{{R_{tot}}}}{r}} \right). \hfill \\ 
\end{gathered}
\vspace{-0.1in}
\end{equation}

From (\ref{eq:Ji_mal}), cost reduction relative to the ideal case is achieved, i.e. $J_i<J_i^0$, when
\begin{equation}
\label{eq:mal_cond}
{\gamma _i}\left| {{D_i}} \right| > \frac{R_{tot}}{r},
\vspace{-0.1in}
\end{equation}
i.e. the $i^{\text{th}}$ user's individual SCR exceeds the average SCR.

By substituting (\ref{eq:Rtot}) into (\ref{eq:mal_cond}) and combining with (\ref{eq:upb}), we obtain the interval of $\gamma_i$ where successful selfish cost manipulation is possible for dishonest user $i$:
\begin{equation}
\label{eq:cond}
\frac{\sigma_i } {{(r - 1)|{D_i}|}}<{\gamma _i}  \leq  \frac{{r{\epsilon ^0} - \sigma_i }}{{|{D_i}|}}.
\end{equation}

If user $i$ is the only dishonest user in the system, i.e. $\sigma_i=0$, then user $i$ must choose a non-negative $\gamma_i \leq {r{\epsilon ^0}}/{{|{D_i}|}}$. However, the knowledge of $\epsilon^0$ is not available in advance, so the user risks ruining cooperation by choosing $\gamma_i$ too large. When there exist other dishonest users, i.e. $\sigma_i>0$, the region in (\ref{eq:cond}) narrows as $\sigma_i$ grows. Moreover, $\sigma_i$ is also unknown to user $i$. Thus, it is very difficult for user $i$ to estimate $\gamma_i$ that satisfies (\ref{eq:cond}) in the proposed distributed game, and users are likely to choose acting honestly to preserve the advantages of cooperation and the associated cost savings $\epsilon^0$. Similar conditions can be derived for the case when users are both ambivalent and dishonest. We conclude that the distributed game provides resilience to dishonest selfish cost reporting.

In summary, the adjustment factor $\gamma_i$ is the product of the ambivalence and dishonest selfish cost reporting factors of user $i$. While the former is likely to be fixed, possibly jeopardizing successful bargaining, the latter is flexible and very risky to manipulate in a distributed system. Thus, the proposed %predictive and 
distributed cost allocation method facilitates honest cooperation decision process for microgrid users.

\vspace{-0.2in}

\begin{figure}[h!]
    \centering
    \includegraphics[width=0.4\textwidth]{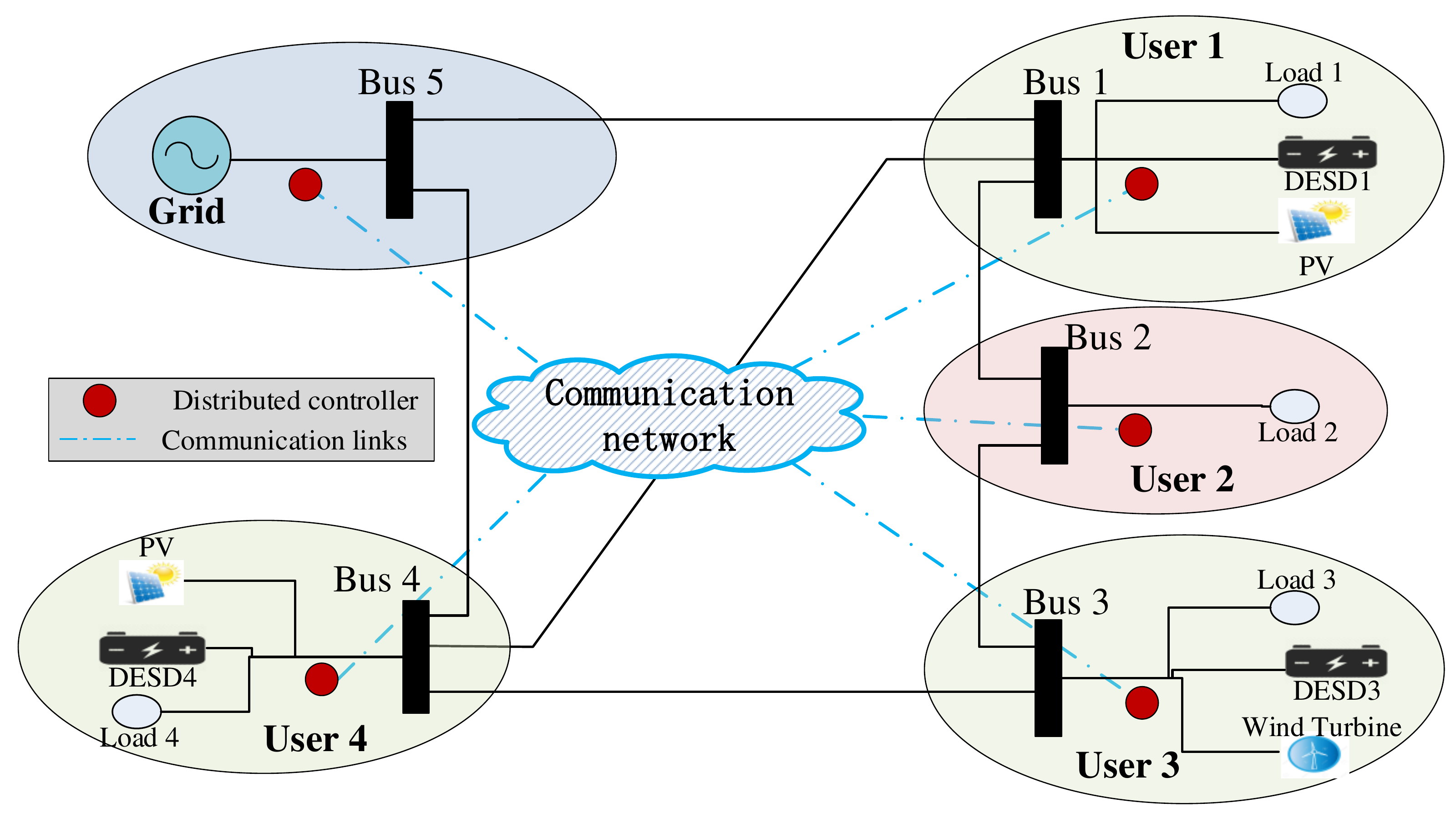}
    \vspace{-0.1in}
    \caption{5-Bus microgrid system}
    \label{fig:5bus}
    \vspace{-0.2in}
\end{figure}
\section{Numerical Results}
In this section, we employ the 5-Bus microgrid system model (4 users and 1 grid) shown in Fig. \ref{fig:5bus} to validate the proposed bargaining game aided by weather-based RG prediction. User 2 is a passive user while Users 1, 3, and 4 are active users, who own DESDs and different types of renewable generators. To protect the users' data, the local information of each Bus, including its RG prediction, power consumption, and battery parameters is only accessible to the controller embedded in that Bus.

\vspace{-0.2in}
\subsection{Data Sets}
\label{sec:prof}
\begin{figure}[h!]
    \centering
    \includegraphics[width=0.4\textwidth]{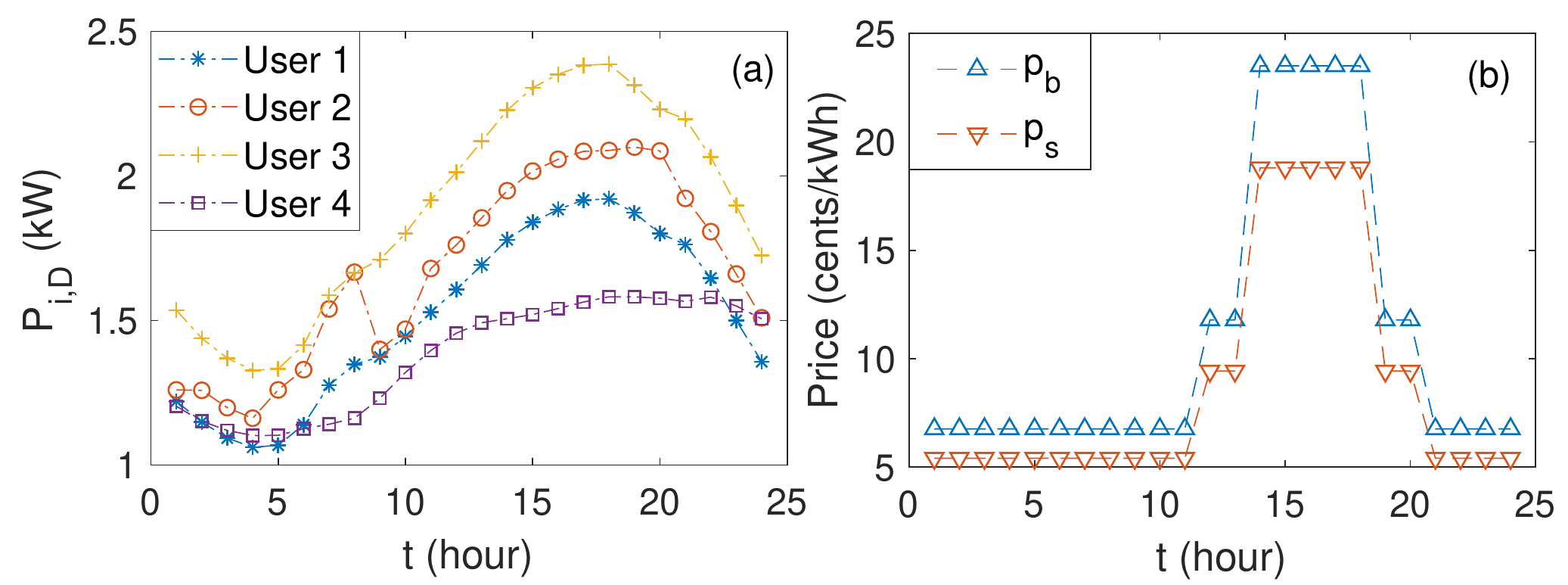}
    \vspace{-0.1in}
    \caption{24-hour (a) demands and (b) prices}
    \label{fig:prof1}
\end{figure}

The following parameters and cases are employed in the numerical results of this section. We investigate day-ahead scheduling and cost allocation, so the interval $T=24$ steps, and $\Delta t=1$ hr. Fig. \ref{fig:prof1} illustrates the demands and prices employed in the simulation over the 24-hour period. Typical demand profiles of all users are obtained from PJM database \cite{PJM}. The buying price $p_b(t)$ is obtained from \cite{TOU}, and the selling price $p_s(t)$ is set to $80\%$ of $p_b(t)$. The active users' DESD and RG parameters are listed in Table \ref{tab:DESD_param}.

\vspace{-0.1in}
\vspace{-0.2in}
\begin{table}[h!]
\vspace{-0.1in}
\caption{Active users' DESD and RG parameters}
\vspace{-0.1in}
\centering
\begin{tabular}{|c|c|c|c|c|c|c|c|}
\hline
User \# & $E_{i,B}^0$ & $E_{i,B}^{min}$ & $E_{i,B}^{max}$ & $P_{i,B}^{max}$ & $\kappa_i$ & RG Type/Size\\
\hline
1 & 2.8kWh & 2.8kWh & 12kWh & 4.3kW & 0.9 &PV/6.5kW\\
\hline
3 & 2.8kWh & 2.8kWh & 7kWh & 3.3kW & 0.9 &WT/4.17kW\\
\hline
4 & 2.8kWh & 2.8kWh & 10kWh & 4.3kW & 0.9&PV/5.3kW\\
\hline
\end{tabular}
\label{tab:DESD_param}
\vspace{-0.1in}
\end{table}

\vspace{-0.1in}

\begin{table}[h!]
\centering
\vspace{-0.1in}
\caption{Weather Forecast}
\label{tab:wp}
\vspace{-0.1in}
\begin{tabular}{|c|c|c|}
\hline
case \#                 & user types                          & weather forecasting results\\ \hline
\multirow{2}{*}{W1} &{PV}& $\{\pi^{\omega_1},\pi^{\omega_2},\pi^{\omega_3}\} = \{0.8,0.2,0\}$ \\  \cline{2-3}
                        & {WT} & $\{\pi^{l_1},\pi^{l_2},\pi^{l_3},\pi^{l_4}\} = \{0,0.3,0.7,0\}$          \\  \hline
\multirow{2}{*}{W2} & {PV}& $\{\pi^{\omega_1},\pi^{\omega_2},\pi^{\omega_3}\} = \{0,0.2,0.8\}$ \\  \cline{2-3}
                    &{WT} & $\{\pi^{l_1},\pi^{l_2},\pi^{l_3},\pi^{l_4}\} = \{0.5,0.5,0,0\}$                    \\ \hline
\end{tabular}
\vspace{-0.1in}
\end{table}

\begin{figure}[h!]
    \centering
    \vspace{-0.1in}
    \includegraphics[width=0.45\textwidth]{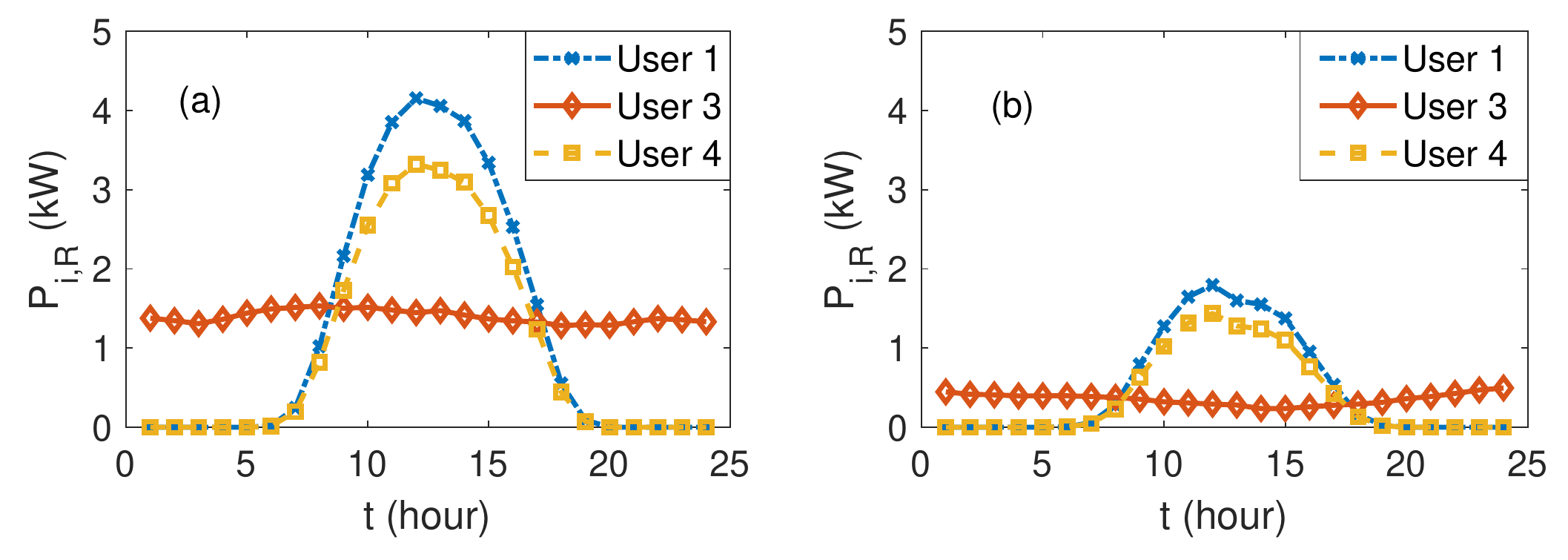}
    \vspace{-0.2in}
    \caption{RG prediction $P_{i,R}(t)$ (\ref{eq:wbr}), $i=1,3,4$: (a) Case W1; (b) Case W2}
    \label{fig:renew_p}
\end{figure}

The RG data for the scenario pool (Fig. \ref{fig:WBR}) are obtained using \cite{SAM} using the weather files for Raleigh, NC, for the solar and the NC Eastern Ocean for the wind RG, respectively. Each user's scenario pool contains 365 scenarios, i.e. $s_{i,k} \text{ for } k=1,\cdots, 365, i=1,3,4$ (Fig. \ref{fig:5bus}) and the respective RG profiles $P_R^{s_{i,k}}(t), \text{ for } t=1,\cdots,24$. For the PV-panel users 1 and 4, the scenarios in their pools are divided into three equal-sized and non-overlapping subsets, based on the daily average sun beam irradiance: sunny $\omega_1$, cloudy $\omega_2$, and rainy $\omega_3$. Within each subset, we assume that the probabilities of the scenarios are uniformly distributed, i.e. $\pi^{s_{i,k}|\omega_m}\sim U[0,1]$. Similarly, for the WT user 3, the scenarios in its pool are divided into four equal-sized and disjoint subsets, based on the daily average wind speed level: $l_1,\cdots, l_4$, where the higher subset index corresponds to a higher wind speed, and $\pi^{s_{i,k}|l_n}\sim U[0,1]$.

Finally, we adopt the weather forecasts for the two cases in Table \ref{tab:wp}. In W1, the weather is likely to be more sunny and windy than in W2, facilitating RG. Using the parameters above, the active users' renewable power generation is predicted by (\ref{eq:wbr}). Fig. \ref{fig:renew_p} shows the prediction results for both weather cases. Unlike for the deterministic renewable generation employed in \cite{AnPES17}, the active users' RG prediction depends on the weather forecast, with significantly higher expected RG power in W1 than in W2.

\vspace{-0.2in}
\subsection{Optimal Power Schedule}
First, we examine the impact of weather-based RG prediction (\ref{eq:wbr}) on the power scheduling results (\ref{eq:P}) using the CoDES method. Once the optimal power schedule ${\mathbf{P}}^*(t)$ is computed distributively, the grid and the active users calculate the power trading cost ${C_{opt}^\mathcal{P}}$ and the individual BDCs ${C_{opt}^{i,\mathcal{B}}}$, respectively.
\vspace{-0.2in}
\begin{figure}[h!]
    \centering
    \includegraphics[width=0.45\textwidth]{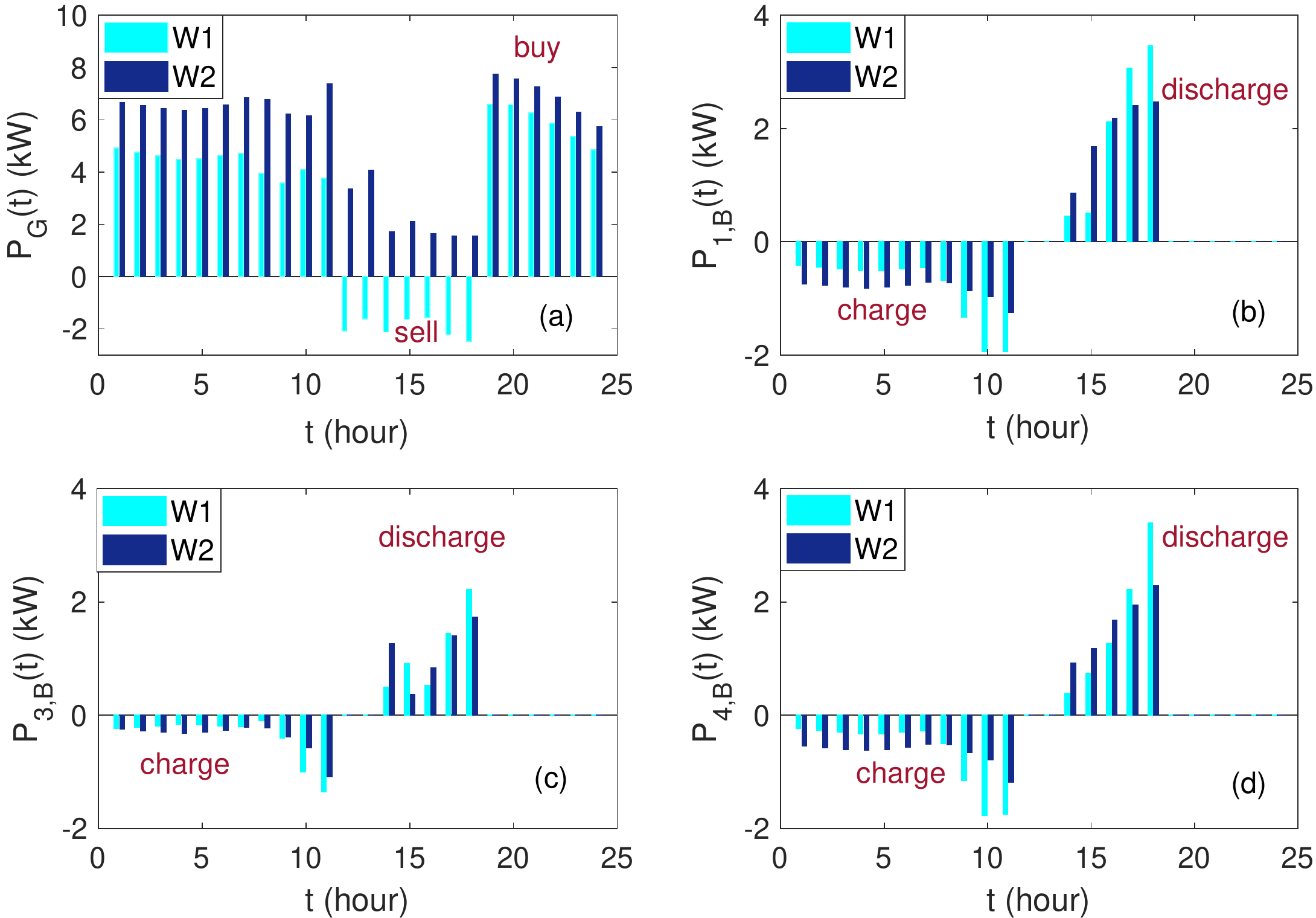}
    \vspace{-0.1in}
    \caption{Power commands for cases W1 and W2: (a) $P_G(t)$ for the grid; (b) $P_{1,B} (t)$ for DESD 1; (c) $P_{3,B} (t)$ for DESD 3; (d) $P_{4,B} (t)$ for DESD 4}
    \label{fig:ps}
    \vspace{-0.1in}
\end{figure}

Fig. \ref{fig:ps} shows the optimal day-ahead power schedule obtained from the CoDES algorithm for weather cases W1 and W2 in Table \ref{tab:wp}. In general, during the off-peak time, when the price is relatively low (Fig. \ref{fig:prof1}), the users tend to buy power and charge the DESDs while they use the stored power in DESDs (discharge) or sell surplus generated power to the grid during the peak usage time, when the utility price is higher. In W2, the total RG is lower than in W1 due to poor weather conditions for RG. Thus, during the time slots $t= 12 \sim 18$ hr, the users have extra generated power to sell back to the grid in W1, but not in W2. In W2, the active users try to buy more power from the grid and store more energy into the DESDs when the utility price is lower, so they can use the stored energy and purchase less power during the peak usage time to reduce the total bill. From the Fig. \ref{fig:ps} (b), we observe that in W2, DESD1 has to charge more power than it does in W1 during the off-peak usage time due to lower RG. It then discharges this stored power during the peak usage time to meet its demand, while in W1, DESD1 discharges to both satisfy the demand and to sell the surplus power back to the grid (Fig. \ref{fig:ps} (a)). The power commands for DESD2 and DESD3 follow the pattern of DESD1 because they cooperate to optimize their power schedules to reduce the overall bill.

Aided by weather-based RG prediction, the power scheduling results are adjusted according to the weather forecast and thus are more realistic than in \cite{AnPES17, stochasticES2014}, and other power scheduling papers. Moreover, the power scheduling optimization problem (\ref{eq:min_J}) takes into account the BDC, which was ignored in \cite{AnPES17}, resulting in significant rearrangement of the DESDs' discharging/charging scheduling commands to reduce the BDC. Note that in Fig. \ref{fig:ps}, due to the BDC optimization, the DESDs actually stop charging or discharging during time slots $12 \sim 13$ and $19 \sim 24$, so all energy exchange during these time slots is among the microgrid users and the main grid. Moreover, considering BDC helps limit the DESDs' State of Charge (SOC) in a safe range ($20\%\sim 100\%$), which aids the longevity of DESDs \cite{7744978}.

To verify the optimality of the CoDES algorithm for the proposed objective function, we also solved the optimization problem (\ref{eq:min_J}) using MATLAB built-in nonlinear programming function \textit{fmincon}. We found that the CoDES algorithm converged to the global optimum found by this centralized algorithm in $\sim 2500$ iterations (6.5 sec), thus confirming the optimality and fast convergence of the CoDES algorithm.

\begin{table}[h!]
\centering
\vspace{-0.1in}
\caption{Ideal cost allocation (in \cent) for W1 and W2}
\label{tab:ica}
\vspace{-0.1in}
\begin{tabular}{|c|c|c|c||c|c|c|}
\hline
\multirow{2}{*}{User \#} & \multicolumn{3}{c||}{W1}                     & \multicolumn{3}{c|}{W2}                      \\ \cline{2-7} 
                        & $J_{soc}$                    & $J_i^0$     & $J_i^0/J_{soc}$  & $J_{soc}$ & $J_i^0$     & $J_i^0/J_{soc}$  \\ \hline
1                       & \multirow{4}{*}{438.68} & -76.16 & -17.36\% & \multirow{4}{*}{1152.87} & 156.55 & 13.58\%  \\ \cline{1-1} \cline{3-4} \cline{6-7} 
2                       &                         & 466.36 & 106.31\% &                          & 472.81 & 41.01\%  \\ \cline{1-1} \cline{3-4} \cline{6-7} 
3                       &                         & 86.65  & 19.75\%  &                          & 373.82 & 32.43\%  \\ \cline{1-1} \cline{3-4} \cline{6-7} 
4                       &                         & -38.16 & -8.7\%   &                          & 149.67 & 12.98\%  \\ \hline\hline
& $\sum_i{D_i}$                   & $D_i$     & $\epsilon^0$ & $\sum_i{D_i}$                   & $D_i$     & $\epsilon^0$ \\ \hline
1                       & \multirow{4}{*}{497.99} & -61.33 & \multirow{4}{*}{14.83}    & \multirow{4}{*}{1186.35} & 164.92 & \multirow{4}{*}{8.37}     \\ \cline{1-1} \cline{3-3} \cline{6-6} 
2                       &                         & 481.18 &     &                          & 481.18 &      \\ \cline{1-1} \cline{3-3} \cline{6-6} 
3                       &                         & 101.48 &     &                          & 382.19 &  \\ \cline{1-1} \cline{3-3} \cline{6-6} 
4                       &                         & -23.34 &     &                          & 158.04 &  \\ \hline
\end{tabular}
\end{table}

\vspace{-0.3in}
\subsection{Ideal Cost Allocation}
\label{sec:ca_h}
In Table \ref{tab:ica}, we illustrate the ideal cost allocation where the selfish costs are given by the individual power costs  $D_i$ (\ref{eq:Di}). The result demonstrates that (\ref{eq:JlessD}) holds, confirming successful bargaining in the ideal case. Obviously, the total bill in W1 is lower than that in W2 while the ideal discount $\epsilon^0$ (\ref{eq:eps0}) on a favorable weather day W1 is greater than that in W2, when weather conditions prevent significant RG. We observe that in both cases, the passive user (User 2) pays the most, followed by User 3, who has the lowest DESD capacity (Table \ref{tab:DESD_param}) and a WT with relatively low RG (Fig. \ref{fig:renew_p}). On the other hand, Users 1 and 4 generate sufficient renewable power to satisfy their demand and/or sell back to the grid, so they are refunded in W1 and have relatively low allocated costs in W2. Finally, the ratios $J_i/J_{soc}$ of active users 1, 3, 4 are lower in W1 than those in W2, confirming that in favorable weather (W1), cooperation is more attractive to active users.

Finally, the distributed NBS cost allocation method in Sec.IV.B converged in 10 iterations ($<1$ sec). This consensus-based algorithm (\ref{eq:xi_up}) converges much faster than the social optimization (\ref{eq:min_J}) because the former's only purpose is to find the average (\ref{eq:x_avg}) while the CoDES algorithm also optimizes the power schedules of the users.

\vspace{-0.1in}
\subsection{Selfish Cost Adjustment and Resilience}
\begin{figure}[h!]
    \centering
    \includegraphics[width=0.47\textwidth]{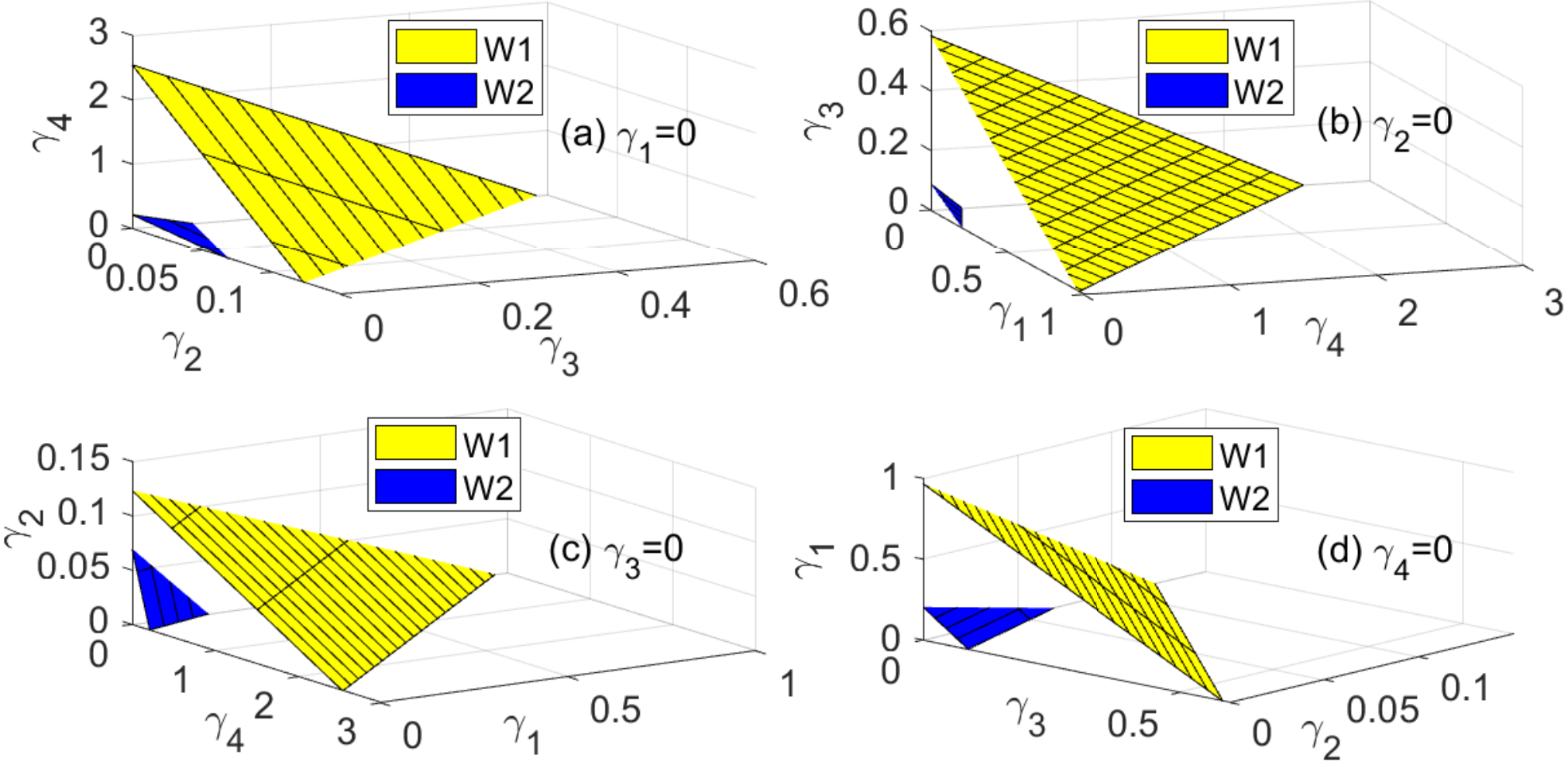}
    \caption{Bargaining is successful for $\{\gamma_i\}$ values below the boundary (\ref{eq:gi_cond}) for W1 and W2 }
    \label{fig:g3d}
    \vspace{-0.1in}
\end{figure}

First, we investigate the effect of the users' adjustment factors $\gamma_i$ on the cooperation outcome (\ref{eq:gi_cond}). Fig. \ref{fig:g3d} shows the boundaries below which successful bargaining is possible as the 3 users' adjustment factors vary and $\gamma_i=0$ for the remaining user. For example, in Fig. \ref{fig:g3d} (a), we assume User 1 is willing to cooperate and is honest (i.e. $\gamma_1=0$), and all combinations of the adjustment factors of Users 2, 3, 4 under the yellow triangle will result in successful bargaining in W1. Note that in W2, the successful bargaining region under the blue triangle is much smaller due to significantly reduced cooperation discount relative to W1 (see Table \ref{tab:ica}). We conclude that when weather forecast is unfavorable for RG, cooperation is less beneficial and the bargaining outcome is more vulnerable to selfish cost adjustment by individual users. Note that the latter conclusion can also be made for any scenario where cooperation does not significantly outperform individual power trading, e.g. for homogeneous users, when either all users are passive or all have large storage capacity and RG resources.
 
Moreover, we observe that successful bargaining outcome is more robust to selfish cost reduction of Users 1 and 4 than of the other two users. For example, in W1, when $\sigma_i=0$, i.e. only user $i$ adjusts its selfish cost (\ref{eq:upb}), bargaining is successful when $\gamma_i\leq 0.9671, 0.1233, 0.5845, 2.5417$, for $i=1,2,3,4$, respectively, corresponding to the vertices of the yellow triangles in Fig. \ref{fig:g3d}, which implies that Users 1 and 4 can reduce their selfish cost $S_i$ significantly below $D_i$ and thus reduce/increase their allocated costs/profits $J_i$ (\ref{eq:newJi}) while still preserving cooperation. In contrast, Users 2 and 3 have less room for cost reduction below $D_i$ without ruining cooperation. This phenomenon is due to the lower individual power cost $|D_i|$ for $i=1,4$ than for other users (see Table \ref{tab:ica}). Thus, when users with high individual costs/profits (large $|D_i|$), who dominate cost allocation, are ambivalent about cooperation, bargaining is more likely to fail than for less dominant users.

\begin{figure}[h!]
    \centering
    \includegraphics[width=0.47\textwidth]{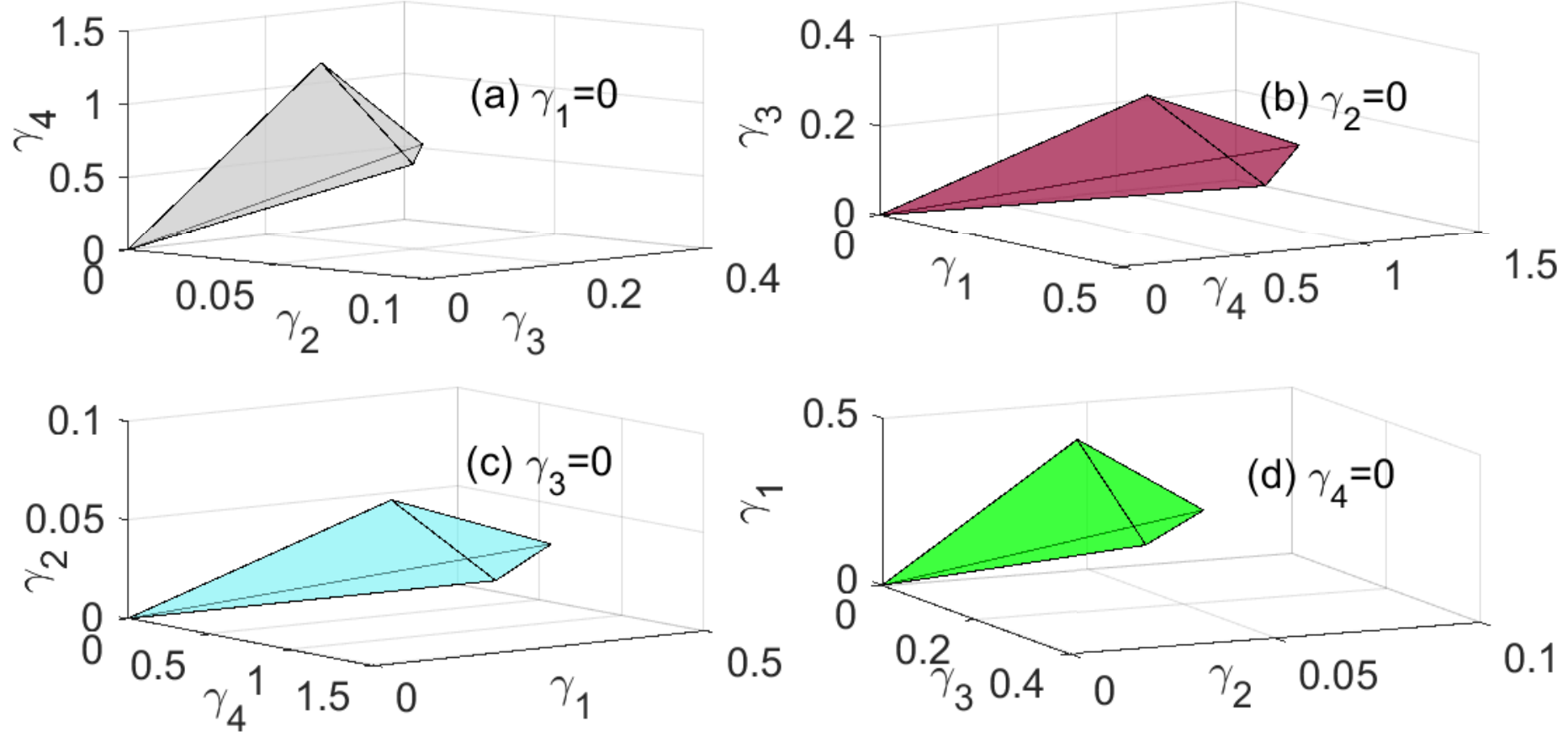}
    \caption{Dishonest users are successful for $\{\gamma_j\}$ values within the shaded regions (\ref{eq:cond}) for W1} 
    \label{fig:g3dc}
    \vspace{-0.3in}
\end{figure}

Whereas above we studied the impact of varying $\gamma_i$ on the bargaining outcome, we now investigate feasibility of malicious cost manipulation. In this analysis, we assume all users are fully interested in cooperation (not ambivalent), but some are dishonest, so $\gamma_j>0$ implies user $j$ reduces its selfish cost to achieve economic benefit. This user is successful if (\ref{eq:cond}) holds. Fig. \ref{fig:g3dc} shows the regions of $\gamma_j$ values that satisfy (\ref{eq:cond}) in W1, and in each subfigure we assume only user $i$ is honest, i.e. $\gamma_i=0$. While Users 1 and 4 (with smaller $|D_i|$ values) have greater flexibility for cost reduction than Users 2 and 3 by satisfying the r.h.s of (\ref{eq:cond}), they are more likely to lose money by violating the l.h.s of (\ref{eq:cond}) if there exist other dishonest users in the system. Thus, the two bounds in (\ref{eq:cond}) greatly constraint feasible regions (Fig. \ref{fig:g3dc}) for successful selfish cost manipulation in W1. In W2, such regions are even smaller due to the lower cooperation discount $r\epsilon^0$.

Satisfying such limited feasibility constraints (\ref{eq:cond}) requires precise knowledge of $r\epsilon^0$ and other users' cost parameters (or $\sigma_i$). However, in the proposed distributed game, this knowledge is not available, and by acting dishonestly, users risk ruining the cooperation or actually losing money. For example, if each dishonest user $j$ selects $\gamma_j$ independently, according to the uniform distribution in $[0,1]$, the probabilities of their choices falling within the shaded region in Fig. \ref{fig:g3dc}(a) and \ref{fig:g3dc}(b) are $0.18\%$ and $1.44\%$, respectively. Even if the dishonest users succeed in choosing their $\gamma_j$ values in the shaded regions, the maximum discount a user can obtain by data manipulation in this region is $J_j^0-J_j =\epsilon^0 =14.83\cent$. This case is on the boundary of the shaded regions in Fig. \ref{fig:g3dc}, where only one dishonest user benefits from data manipulation and the only honest user $i$ pays its selfish cost $D_i$. The average benefit $J_j^0-J_j$ per dishonest user does not exceed $\epsilon^0/3 = 4.94\cent$ within the shaded region.

A much more likely outcome is that dishonest users' $\gamma_j$ values fall outside the shaded regions in Fig. \ref{fig:g3dc}. The probabilities of this event are 99.82\% in Fig. \ref{fig:g3dc}(a) and 98.56\% in Fig. \ref{fig:g3dc}(b), respectively. First, consider the case when the upper bound in (\ref{eq:cond}) is satisfied for all users, i.e. bargaining is successful, but the lower bound in (\ref{eq:cond}) does not hold for one or more dishonest users. The probability of this event is 2.19\% in Fig. \ref{fig:g3dc}(a) and 17.2\% in Fig. \ref{fig:g3dc}(b), respectively. In this case, the latter dishonest users obtain a lower discount than if they acted honestly. Thus, their malicious attempts are not successful. Finally, the most likely outcome of selfish cost adjustment is unsuccessful bargaining, which has the probability 97.63\% in Fig. \ref{fig:g3dc}(a) and 81.4\% in Fig. \ref{fig:g3dc}(b), respectively. This event occurs when the upper bound in (\ref{eq:cond}) does not hold and corresponds to the region above the yellow triangle in Fig. \ref{fig:g3d}, resulting in the loss of $\epsilon^0 = 14.83\cent$ for all users relative to the case when all users act honestly. Note that the magnitude of this loss significantly exceeds expected profit of a successful malicious user. Note that the likelihoods discussed above are similar to those in Fig. \ref{fig:g3dc}(c,d), and the chances of successful selfish cost misrepresentation decrease with the number of malicious users. These examples demonstrate that in the absence of knowledge of all other users' selfish costs as well as the social cost $J_{soc}$, malicious users are very likely to lose money. Thus, it makes sense for them to act honestly. Therefore, we conclude that the proposed game provides resilience to malicious user behavior.

\vspace{-0.1in}
\section{Conclusion}
We proposed a fully distributed bargaining game for cooperative energy management in microgrids that includes a multi-step scheduling algorithm (CoDES) and NBS-based cost allocation method. A weather-based stochastic RG prediction method was incorporated in the proposed game. Moreover, the impacts of users' ambivalence about cooperation and dishonest behavior on the bargaining outcome were analyzed. For a grid-connected microgrid with 4 heterogeneous users, it is demonstrated that the overall day-ahead predicted cost of power purchased from the grid and vulnerability to users' selfish cost adjustment decrease while the cooperation discount increases as weather forecast becomes more favorable for renewable power generation. Moreover, bargaining is more likely to fail when users with dominant selfish costs/profits are ambivalent or dishonest. Finally, it was demonstrated that the proposed game is resilient to dishonest selfish cost reporting.


%% use section* for acknowledgment
%\section*{Acknowledgment}
\vspace{-0.1in}

\bibliographystyle{ieeetr}
\bibliography{ref_all}

\begin{IEEEbiography}
[{\includegraphics[width=1in,height=1.25in,clip,keepaspectratio]{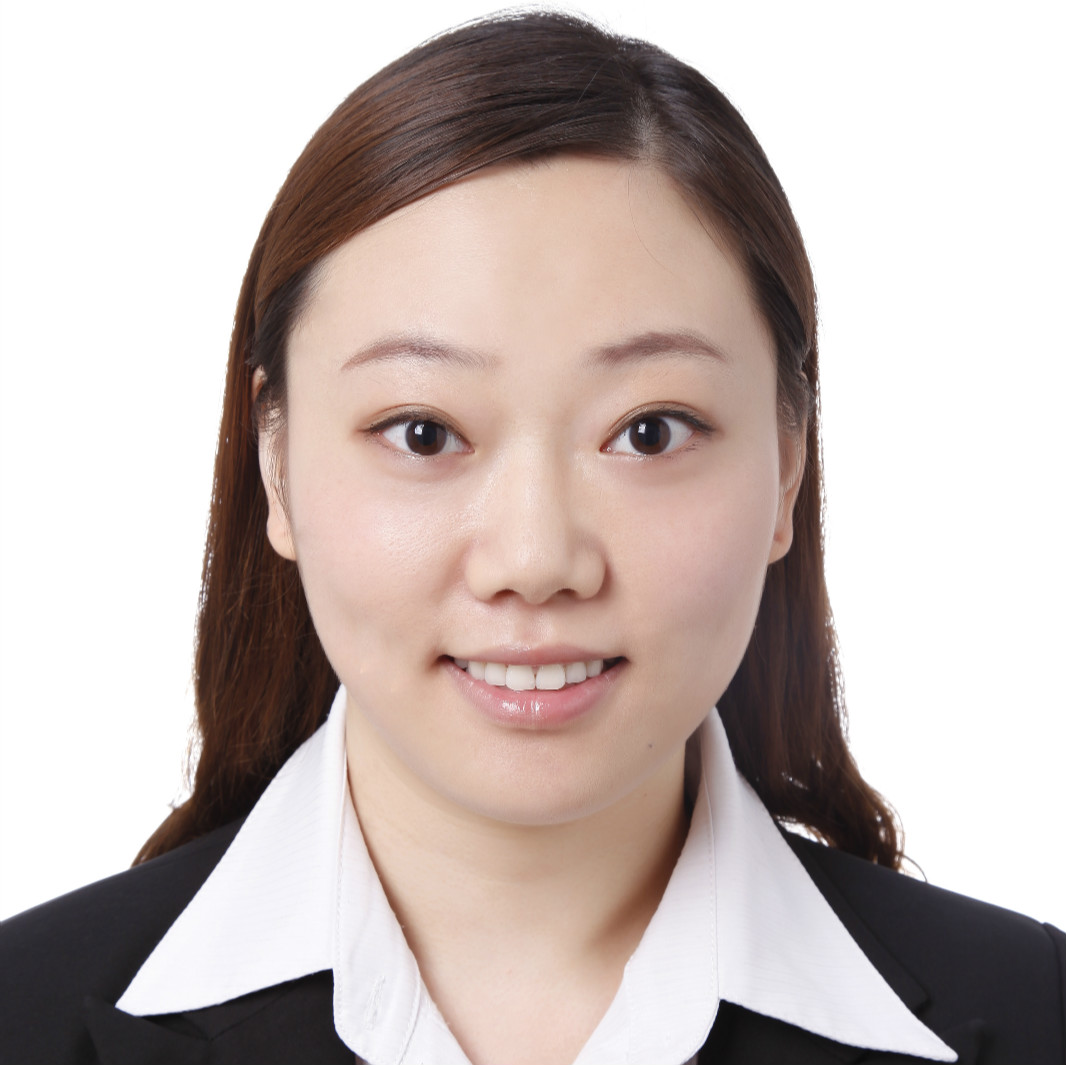}}]{Lu An}
received her B.S. (2012) and M.Eng. (2015) in Electrical Engineering from Beijing University of Posts and Telecommunications, China. During her master degree study, she was working on energy efficiency optimization in green wireless communication system. She is currently pursuing the Ph.D. degree in Electrical Engineering with the Department of Electrical and Computer Engineering at North Carolina State University, Raleigh, NC, USA. Her research interests include game-theoretic methods in energy management and smart grid security.
\end{IEEEbiography}
\vspace{-0.5in}
\begin{IEEEbiography}[{\includegraphics[width=1in,height=1.25in,clip,keepaspectratio]{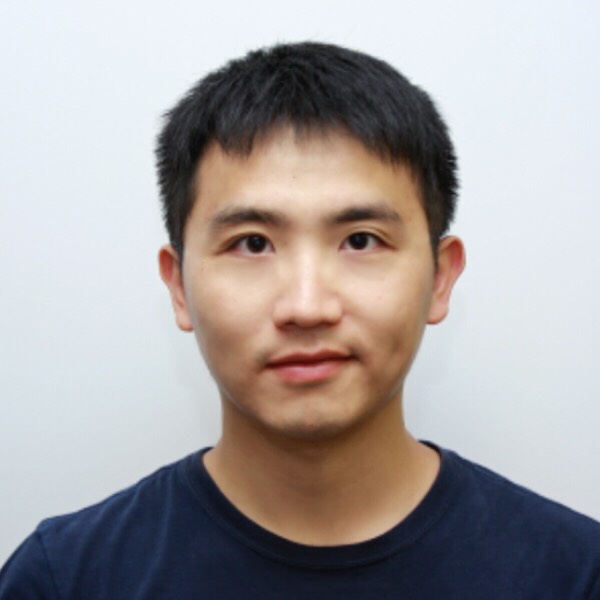}}]{Jie Duan}
(S'14) received his B.S. in Electrical Engineering from Zhejiang University, China, in 2012. In 2014, he received his M.Sc. in Electrical Engineering from Xi'an Jiaotong University, China. He is currently working toward the Ph.D. degree in electrical engineering in the Department of Electrical and Computer Engineering, North Carolina State University, Raleigh. He has been a part of the Advanced Diagnosis, Automation, and Control Laboratory at North Carolina State University since September 2014. 
His research interests include distributed control, cyber-physical security with application in smart grids, and energy management systems. 
\end{IEEEbiography}
\vspace{-0.5in}
\begin{IEEEbiography}[{\includegraphics[width=1in,height=1.25in,clip,keepaspectratio]{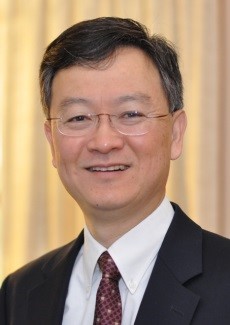}}]{Mo-Yuen Chow}
earned his degree in Electrical and Computer Engineering from the University of Wisconsin-Madison (B.S., 1982); and Cornell University (M. Eng., 1983; Ph.D., 1987). Dr. Chow is a Professor in the Department of Electrical and Computer Engineering at North Carolina State University. Dr. Chow was a Changjiang Scholar and a Qiushi Professor at Zhejiang University.
Dr. Chow’s recent research focuses on distributed control and management on smart grids, batteries, and robotic systems. Dr. Chow has established the Advanced Diagnosis, Automation, and Control Laboratory. He is an IEEE Fellow, the Co-Editor-in-Chief of IEEE Trans. on Industrial Informatics 2014-2018, Editor-in-Chief of IEEE Transactions on Industrial Electronics 2010-2012. He has received the IEEE Region-3 Joseph M. Biedenbach Outstanding Engineering Educator Award, the IEEE ENCS Outstanding Engineering Educator Award, the IEEE ENCS Service Award, the IEEE Industrial Electronics Society Anthony J Hornfeck Service Award. He is a Distinguished Lecturer of IEEE IES.
\end{IEEEbiography}
\vspace{-0.5in}
\begin{IEEEbiography}[{\includegraphics[width=1in,height=1.25in,clip,keepaspectratio]{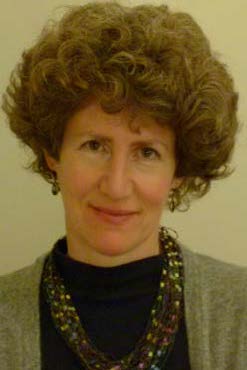}}]{Alexandra Duel-Hallen}
received her Ph.D. in Electrical Engineering from Cornell University in 1987. During 1987-1993, she was with the School of Electrical Engineering, Cornell University, Ithaca, NY and the Mathematical Sciences Research Center, AT\&T Bell Laboratories, Murray Hill, NJ. She joined the Electrical and Computer Engineering Department at North Carolina State University in 1993. She served as an Associate Editor for the IEEE Transactions on Communications and as a Guest Editor for the IEEE Journal on Selected Areas in Communications. She is listed in ``American Men and Women in Science," Cengage Learning and in \textit{Thomson Reuters Highly Cited Research}. Her paper was included in \textit{The Best of the Best: Fifty Years of Communications and Networking Research}, IEEE Press, 2007 and the \textit{IEEE Communications Society 50th Anniversary Journal Collection}. She is a Fellow of IEEE. Her current research interests are in wireless communications and smart grid. 
\end{IEEEbiography}

\end{document}